\documentclass[sigconf]{acmart}

\usepackage[algoruled]{algorithm2e}
\usepackage{amsmath}
\usepackage{balance}
\usepackage{xcolor}
\usepackage{graphicx}
\usepackage{amsthm}
\usepackage{caption}
\usepackage{subcaption}
\usepackage{url}
\usepackage{breakurl}

\usepackage{hyperref}
\hypersetup{colorlinks=true, urlcolor=blue!50!black, linkcolor=blue!50!black, citecolor=blue!50!black}
\usepackage[noabbrev,nameinlink,capitalise]{cleveref}

\renewcommand*{\proofname}{Proof (sketch)}
\newcommand{\textcircledcentred}[1]{\textcircled{\raisebox{-0.8pt}{#1}}}

\newtheorem{guarantee}{Guarantee}
\AtBeginDocument{%
  }

\setcopyright{acmlicensed}
\copyrightyear{2018}
\acmYear{2018}
\acmDOI{XXXXXXX.XXXXXXX}

\acmConference[Conference acronym 'XX]{Make sure to enter the correct
  conference title from your rights confirmation email}{June 03--05,
  2018}{Woodstock, NY}
\acmISBN{978-1-4503-XXXX-X/18/06}

\begin{document}

\title{Defending Against Attack on the Cloned: In-Band Active Man-in-the-Middle
Detection for the Signal Protocol}


\author{Wil Liam Teng}
\affiliation{%
  \institution{University of Oxford}
  \city{Oxford}
  \country{United Kingdom}
}
\email{wil.teng@cs.ox.ac.uk}

\author{Kasper Rasmussen}
\affiliation{%
  \institution{University of Oxford}
  \city{Oxford}
  \country{United Kingdom}
}
\email{kasper.rasmussen@cs.ox.ac.uk}


\begin{abstract}
Signal is one of the most popular secure messaging protocols in use
today, and offers user authentication and end-to-end encrypted
messages. However many of the security guarantees, especially when
facing and active MitM attacker, depend on knowing the correct public
key of the other client. Signal offers a way to manually compare keys
if two users are physically next to each other but few users do this
on a regular basis. In fact, studies~\cite{usabilityBlog,
  usabilityUsenix1, usabilityUsenix2, usabilityNdss} have shown that
users in general have very little knowledge of this authentication
mechanism, or of the reasons why it is important for security.

In this paper we propose an extension to the Signal protocol that
automates the process of key confirmation without relying on the
participation of users, and without using an out-of-band communication
channel.  We are able to achieve this by using the server to keep
track of a fingerprint as key ratcheting is performed. This is
convenient and transparent to the users because the server is always
online, but comes at the cost of slightly altered trust assumptions on
the server. We present a detailed description of the protocol, and
document how it differs from Signal as well as the new security
guarantees. We also show that Signal's existing security guarantees
still hold in our modified version. We prove our protocol secure with
respect to a powerful active MitM adversary, that not only controls
the communication channel, but can also corrupt a client, i.e., has
(one time) access to all secrets on the client.  Our proof-of-concept
implementation, is based on the open-source Signal library used in
real-world instant messaging applications. It shows that our solution
is practical and integrates well with the library.  Our performance
measurements show that our protocol has an excellent performance, with
a tiny and insignificant overhead compared to Signal.
\end{abstract}




\maketitle

\section{Introduction}
Instant messaging applications are used extensively for day-to-day communication and have been well-integrated into our everyday lives. In this context, the Signal protocol (or simply, Signal) is widely regarded as the de facto standard for end-to-end encrypted messages, and has been serving as the underlying protocol for popular instant messaging applications such as Signal's own instant messaging application, WhatsApp \cite{whatsapp}, and Facebook Messenger \cite{messenger} where WhatsApp alone has nearly 3 billion users worldwide in 2023 \cite{whatsappStatista}. One ``key" problem (quite literally) in Signal is key authentication, which is to make sure that the messaging keys used to send messages in the protocol actually belongs to a specific user. The current authentication method is for the communicating pair of users to compare the fingerprints of their keys in an out-of-band fashion and this requires the active participation of the users. However, usability studies \cite{usabilityBlog, usabilityUsenix1, usabilityUsenix2, usabilityNdss} have shown that the awareness of users of participating in this procedure is quite low. Furthermore, because the user keys change every time a key-ratcheting step is performed, the users would have to manually authenticate their keys frequently which is unlikely to ever be done in practice. While recent work~\cite{euroSP,cloneIdentityKey,cloneOrdinal} improves on this out-of-band authentication procedure, they ultimately still rely on users' interaction.

At the same time, news of governments coercing companies to disclose user data are getting increasingly common, as shown by the list of government requests received by Signal \cite{signalBigbrother}. The European Union is currently in the process of passing a new legal framework eIDAS \cite{eidas} and Article 45 of eIDAS compels Internet browsers to trust root certificates issued by government chosen certificate authorities \cite{eff, mozilla}. In response, an open letter by global cybersecurity experts \cite{euArticle45OpenLetter} condemns that Article 45 ``radically expands the ability of governments to surveil both their own citizens and residents across the EU by providing them with the technical means to intercept encrypted web traffic". In the United Kingdom, the recently passed Online Safety Act 2023 \cite{ukOnlineSafetyAct} that allows user messages to be monitored \cite{wiredOnlineSafetyAct} has also received criticisms from cybersecurity experts. The open letters condemn the law as being able to ``undermine end-to-end encryption" \cite{onlineSafetyActOpenLetter1} by ``creating a new power to compel online intermediaries to use \emph{accredited technologies} to conduct mass scanning and surveillance of all citizens on private messaging channels'' \cite{onlineSafetyActOpenLetter2}. These policies have the potential domino effect that organisations such as Internet Service Providers and business companies could be lawfully obliged to act as a Man-in-the-Middle (MitM) and break end-to-end encryption to tap into the private communication channels of their users. 

In light of these issues, we turn to the existing Signal protocol and build improvements on the key authentication of the protocol as a novel solution for these extreme situations to automate the process of key authentication and to detect active MitM attacks without relying on the interaction of the users. We go a step further by considering a stronger adversary, such as a malicious organisation, that does not only has the power to control the communication channels and act as an active MitM, but also has the ability to clone the state of a client device, thus gaining access to all the client's secret key material, including both long-term and ephemeral secrets.

Much of the existing work on Signal consider a fully untrusted server. This assumption however is often not entirely true in practice as messaging applications are unusable when client devices are not connected to the server. This means that the server is at a minimum trusted to relay the envelopes between participants and to handle the ``correct'' key distribution for end-to-end encryption to be effective. Our solution makes use of the ever present server to automatically detect the presence of a powerful active MitM in band with the existing communication channels. We do so while also maintaining the existing security properties of Signal, namely, end-to-end encryption, perfect forward secrecy, post-compromise security; as well as confidentiality and integrity of the messages. We summarise our main contributions as follow:

\begin{itemize}
\setlength{\itemsep}{0em}
\item We propose additions to the Signal protocol that make it resistant to \emph{active} MitM attacks (as opposed to only passive) without requiring user intervention. Our solution automatically detects the presence of an active MitM attacker even though he has access to a snapshot of all secrets of a client device, while preserving the existing security properties of Signal.
\item We analyse our new protocol with respect to both the new security properties we provide, and the existing properties the basic Signal protocol provides.
\item We implement a proof-of-concept in Rust of our modified Signal protocol based on the open-source Signal library to demonstrate the practicality of our solution.
\item We run an experiment and measure the performance characteristics to benchmark our solution against the original Signal protocol.
\end{itemize}

\section{Related Work}

First proposed by \cite{signalDocs}, the Signal protocol (or Signal) is based on the Extended Triple Diffie-Hellman (X3DH) key agreement protocol \cite{x3dh} and the Double Ratchet Algorithm \cite{doubleRatchet}. The analysis on Signal has first been formalised in \cite{signalFormal}, and \cite{pcs} coined the term \emph{Post-Compromise Security} (PCS) referring to the protocol's self-healing property. Due to its widespread practical uses, Signal has since attracted other research work \cite{signalUc, doubleRatchetEurocrypt19, doubleRatchetCrypto22, said, healingSpeedTaxonomy}. Our work is largely inspired by research on key authentication and cloning attacks in Signal, and we focus on the related work in these two topics.


\subsection{Signal Key Authentication}

Signal's original key authentication method depends on the pair of users comparing the \emph{Safety Numbers} or the fingerprints of their devices' identity out-of-band \cite{signalSafetyNumber} derived from hashing the identifiers and the long-term public keys of the devices. This process of comparing the fingerprints out-of-band is also known as an authentication ceremony \cite{authCeremony}. As the fingerprints are calculated on purely static information, \cite{euroSP} improves the derivation method of the Safety Numbers to reflect the ratcheted keys. This prevents an adversary who can manipulate the encoded fingerprints from performing an active man-in-the-middle (MitM) attack, but the solution ultimately still depends on the users performing an authentication ceremony. Solutions proposed to alleviate the out-of-band authentication ceremony \cite{whatsappKeyTrans, decim} makes use of a publicly auditable append-only key transparency systems, but this requires additional parties to the Signal communication infrastructure. Motivated by the current events and the usability problems \cite{usabilityBlog, usabilityUsenix1, usabilityUsenix2, usabilityNdss} in out-of-band authentication, we propose a solution that does not rely on user action or additional parties in Signal's communication infrastructure.

\subsection{Cloning Attacks in Signal}

Cloning attacks in the context of Signal have been studied in \cite{cloneCounter, cloneSesame, cloneIdentityKey, cloneOrdinal}. The solution proposed by \cite{cloneCounter} embeds an increasing counter into the protocol as part of the associated data for each envelope sent, whereas \cite{cloneSesame} improves Signal's session management scheme to detect a cloned device by checking if envelopes are encrypted with a key from a session not recognised by an honest device. Both solutions assume a clone operating only through the interface of the messaging applications and not having direct access to the keys. \cite{cloneIdentityKey} proposes a clone detection mechanism that reintroduces the long-term identity keys of clients to generate signatures on past envelopes. The solution assumes an active MitM adversary that have access to the state of the clients, including the long-term keys, but still requires an out-of-band channel for users to detect the compromise of the long-term keys. \cite{cloneOrdinal} associates a ciphertext with an index called an ordinal, and attaches the history of previously sent and received ciphertexts' ordinals (and hashes) to a subsequent ciphertext for the detection of a message forgery by an adversary who can control the communication channels and expose the state of a device (similar to our adversary model). While \cite{cloneOrdinal, cloneCounter, cloneSesame} require a client to keep track of a different state for each of its communication peers, our solution only needs a client to keep track of a single hash regardless of its number of communication peers. Our solution also detects a much more powerful MitM adversary than \cite{cloneCounter,cloneSesame,cloneIdentityKey} who not only has full control of the communication channels but also a one-time direct access to all of a client's secret keys, including its long-term identity keys.

\subsection{Comparison to Existing Work}

In this paper we present our solution that builds directly on top of the Signal protocol without introducing any additional third-party to the communication infrastructure or without any user intervention for the authentication ceremony. Our solution is only applied on the interaction between the server and the client in the protocol, and does not change the interaction between clients or the derivation of the session keys. As such, existing solutions \cite{signalSafetyNumber, euroSP, whatsappKeyTrans, decim, cloneCounter, cloneSesame, cloneIdentityKey, cloneOrdinal} can still be integrated seamlessly with our solution. We thus offer a tradeoff and an alternative trust model for applications building on top of Signal: either trust the Signal server but not the users' environment using our solution for the detection of a more powerful active MitM adversary; or for security-savvy users who do in fact perform authentication ceremonies, to not trust the Signal server at all but only recover security against a passive state-cloning adversary (Signal's adversary model for post-compromise security) \cite{pcs}.

\section{Background}
\label{sec:background}

As our solution is built on top of Signal, we provide an overview of the protocol as background in this section. Signal is used for generating new session keys between two clients, an initiator and a responder, who take turns sending envelopes in an asynchronous messaging environment. Based on the Triple Diffie-Hellman (X3DH) handshake and the Double Ratchet Algorithm, Signal provides important security guarantees, i.e., content confidentiality and integrity, authentication, perfect forward secrecy, and post-compromise security. 

\subsection{Phases of the Signal Protocol}

Signal is further divided into four subprotocols, one for each phase of communication, i.e., the registration phase, the session establishment phase, the asymmetric ratcheting phase, and the symmetric ratcheting phase. The registration phase is for a client to generate and commit its \emph{prekey bundle} (explained later in this section) to the server. In the session establishment phase, an initiator establishes a communication session with a responder, by retrieving the prekey bundle of the responder from the server and computing the session keys to generate the first \emph{envelope} (details in \cref{sec:messSending}) containing the encrypted content. After the initiator sends the first envelope to the responder and the responder has received the envelope and decrypted the content, the initiator and the responder have entered the first epoch of the communication session. The communication session enters a new epoch if the roles of the sender and the recipient of the envelopes are swapped, and this occurs in the asymmetric ratcheting phase. In this phase, the sender generates new keying materials to update (or \emph{ratchet}) the session keys and sends the keying materials along with the first envelope of the new epoch to the recipient. In the same epoch, i.e., when the roles of the sender and the recipient do not change, if the sender sends additional consecutive envelopes after the first envelope of that epoch, the symmetric ratcheting phase takes place. Session keys are ``rolled over'' in this phase without introducing new keying materials for each consecutive envelope, and the chain of envelopes is known as the message chain. 

\subsection{Algorithms in the Signal Protocol}

We encapsulate the detailed steps of the four phases into algorithms which we use later to describe our improved Signal protocol, and we list the algorithms here. We provide the detailed steps of the algorithms in pseudocode in \cref{app:algorithms}, and explain only their relevant details in this subsection. Note that all public-private keypairs generated in the algorithms are in the form of Diffie-Hellman keypairs. We denote an epoch as $t$, and the number of envelopes sent or received by a client as $n$. We collectively refer to the secrets that a client stores as state, denoted $s^{t,n}$ for the state in an epoch $t$ after the processing of envelope $n$. Also note that the state of the client is updated accordingly after the execution of the algorithms. 

$(pkb, s^{0, 0}) \leftarrow \mathtt{PkbGen}(\lambda)$: Prekey bundle generation (\cref{alg:pkbGen}). This algorithm generates the prekey bundle of a client and initialise the client's state. The algorithm takes in a security parameter $\lambda$ as input and gives out two tuples, namely, the prekey bundle $pkb$ and an initial state $s^{0, 0}$. The prekey bundle consists of three public keys called the identity key $g^{x_{i}}$, signed prekey $g^{x_{s}}$, and a one-time prekey $g^{x_{o}}$ respectively. Note that in Signal multiple one-time prekeys can be included in the prekey bundle but for simplicity we assume only a single one-time prekey for each client in this paper. The prekey bundle also includes a signature on $g^{x_{s}}$ using the private identity key $x_{i}$, i.e., $Sign_{x_{i}}(g^{x_{s}})$. The initial state consists of the three corresponding private keys of the public keys in the prekey bundle, that is, $(x_{i}, x_{s}, x_{o})$. 

$(epk^{t+1}, s^{t+1, 0}) \leftarrow \mathtt{EphKeyGen}(\lambda, s^{t,n}, phase)$: Ephemeral key generation (\cref{alg:ephKeyGen}). This algorithm generates the random ephemeral keypairs, denoted $(epk^{t+1}, esk^{t+1})$, required for an initiator or a sender to respectively establish session keys in session establishment or update the session keys in asymmetric ratcheting. Both cases update the state of the client accordingly. The algorithm requires a security parameter $\lambda$, the client's current state $s^{t,n}$, and the phase of the communication session as inputs. The algorithm returns a tuple $epk^{t+1}$ containing the freshly generated public keys for the new epoch $t+1$ depending on the communication phase, and the client's updated state $s^{t+1, 0}$ as outputs. The updated state includes $esk^{t+1}$, a tuple containing the corresponding freshly generated private keys of $epk^{t+1}$. If the phase is session establishment, $epk^{t+1}$ consists of two public keys, a base key $g^{x_{b}}$ and the first ratchet key $g^{x^1}$; otherwise if the phase is asymmetric ratcheting, $epk^{t+1}$ contains a single ratchet key $g^{x^{t+1}}$. Note that this algorithm is not executed for the recipient of an envelope.

$(s^{t,n+1}) \leftarrow \mathtt{MessKeyGen}(s^{t,n}, phase, pk_B)$: Message key generation (\cref{alg:messKeyGen}). The algorithm establishes the session keys according to the X3DH algorithm if the phase is session establishment, or updates the session keys according to the Double Ratchet algorithm if the phase is asymmetric ratcheting or symmetric ratcheting. The session keys consist of three symmetric keys: the root key $rk$, the chain key $ck$, and the message key $mk$. The chain key and the message key are derived from the root key, and they are updated for every new envelope. The root key is updated on every asymmetric ratcheting. The algorithm accepts three inputs, the current state $s^{t,n}$ of the client, the phase of the communication session, and the public key material $pk_B$ of a communicating partner $B$. The algorithm returns one output: the updated state $s^{t,n+1}$ that includes the established or updated session keys. Depending on the phase, previous session keys are removed from the state.

\subsection{Sequence of Algorithm Execution}

\begin{figure}
\centering
\includegraphics[width=\linewidth]{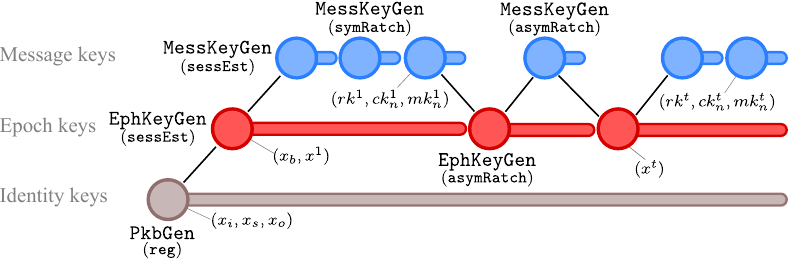}
\caption{Sequence of algorithm execution throughout the four communication phases in the Signal protocol, namely, registration ($\mathtt{reg}$), session establishment ($\mathtt{sessEst}$), symmetric ratcheting ($\mathtt{symRatch}$), and asymmetric ratcheting ($\mathtt{asymRatch}$). Each of the three algorithms, namely, $\mathtt{PkbGen}$, $\mathtt{EphKeyGen}$, and $\mathtt{MessKeyGen}$, changes the state of a sender represented by the tuple depending on the phase of the communication.}
\label{fig:stateTrace}
\end{figure}

We now provide a trace in \cref{fig:stateTrace} on the sequence of algorithm execution by a client throughout the four phases of Signal. We refer to the previous subsection for more details on the algorithms and the content of the state. 

The state of a client is first initialised by $\mathtt{PkbGen}$ in the registration phase ($\mathtt{reg}$). To initiate a conversation in the session establishment phase ($\mathtt{sessEst}$), an initiator runs $\mathtt{EphKeyGen}$ to generate the required keying materials and then $\mathtt{MessKeyGen}$ to establish the initial session keys for the first envelope in the first epoch. The initiator can then choose to generate and send more envelopes without waiting for a reply from the responder after the first envelope in the same epoch. Generating and sending the second envelope onwards indicate that the initiator enters the symmetric ratcheting phase ($\mathtt{symRatch}$) where the initiator runs $\mathtt{MessKeyGen}$ to update the session keys for each successive envelope. The communication session is then continued indefinitely with the initiator and the responder taking turns acting as the sender. If a role switch occurs, we say that the asymmetric ratcheting phase ($\mathtt{asymRatch}$) takes place where the sender runs both $\mathtt{EphKeyGen}$ and $\mathtt{MessKeyGen}$ in sequence to update the session keys for the first envelope of the new epoch. Similarly the sender enters the symmetric ratcheting phase if there are multiple subsequent envelopes in that same epoch.

Note that the registration phase is executed once in the lifetime of a client and the session establishment phase is executed once for each communicating partner. Also observe that after session establishment, the communication phase alternates between the asymmetric ratcheting phase and the symmetric ratcheting phase. In \cref{fig:stateTrace} some secrets are removed or replaced from the state depending on the algorithms executed and the communication phase, details of which are covered in \cref{app:algorithms}.

\section{System and Adversary Model}
\label{sec:sysAdvModel}

\begin{figure}[!t]
\centering
\includegraphics[scale = 1]{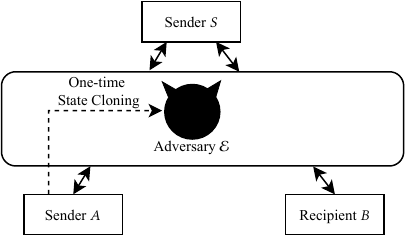}
\caption{Our system and adversary model. The system model is identical to that of Signal's. It consists of three parties, a sender $A$ and a recipient $B$ communicating via an honest server $S$ that forwards envelopes. The adversary model consists of adversary $\mathcal{E}$ capable of manipulating the messages on the communication channels between $A$ and $S$, and between $S$ and $B$. The adversary has the additional ability to clone the state of $A$ at most one time, thereby accessing all $A$'s secrets.}
\label{fig:sysAdvModel}
\end{figure}

In this section we present our system model and adversary model depicted in \cref{fig:sysAdvModel}. These include our assumptions on the system participants as well as the adversary capabilities and its goals. We later prove that the adversary's goals are not achieved according to the guarantees of our protocols.

\subsection{System Model}

Our system model is identical to the underlying communication model for Signal. The system model consists of three parties: a sender $A$, a recipient $B$, and a server $S$. An envelope sent from the sender to the recipient is always sent via the server, unless dropped by an adversary. While Signal is often presented as if the server is fully untrusted, we argue that in practice the Signal server is in fact semi-trusted, since it is expected to forward messages between participants and distribute the correct keys, i.e., not act as a Man-in-the-Middle (MitM), in order to guarantee end-to-end encryption. We assume that the server correctly distributes client keys, computes the required hash values and takes action to end the connection if verification fails (as described in \cref{sec:protocolDesc}). Beyond this there are no trust assumptions on the server, specifically the server will never be able to access un-encrypted envelopes.

We assume that the communication channels between the clients and the server are protected with Transport Layer Security (TLS) \cite{tls} (as is often the case in practice). The server is always online, whereas the sender and the recipient need not be. If the recipient is offline, the server stores the envelope until the envelope is fetched by the recipient. The sender and the recipient are required to be online when sending an envelope, or when fetching an envelope from the server. We further assume that the sender and the recipient can securely delete the ephemeral secrets when such deletions are required as specified by the Signal protocol.

\subsection{Adversary Model}
 
We consider a powerful adversary $\mathcal{E}$ capable of acting as a MitM to observe the messages exchanged in the protocol. We assume that the adversary has full control of both channels, i.e., the channel between a sender and the server, and the channel between the server and a recipient. We assume that the underlying cryptographic primitives are secure, and that the adversary acts as a Dolev-Yao \cite{dolevYao} attacker, meaning that the adversary can replay, drop, and modify messages on the two channels. 

We allow the adversary to act any time after the end of the Registration Protocol.  We give the adversary the ability to clone the state of the sender at most once, i.e., without modifying the existing state on the sender. This means that the adversary has access to sender $A$'s secrets at the epoch of compromise, which we denote $\tau$. Specifically, we allow the adversary to obtain the tuple $(x_{i_A}, x_{s_A}, x_{o_A}, x_{b_A}, x_A^{\tau}, rk^{\tau}, ck^{\tau}_n, mk^{\tau}_n)$, details of which are in \cref{sec:background}. Recall that the communication enters a new epoch whenever an asymmetric ratcheting is completed, i.e., whenever the roles of the sender and the recipient are switched between the initiator and responder.

The four goals of the adversary are: (1)~to remain undetected in the system after sending envelopes impersonating the sender, (2)~to retrieve the content of past envelopes in epochs before $\tau$, (3)~to retrieve content of future envelopes sent by the sender $A$ after $\tau$ when the sender is eventually online, and (4)~to break the confidentiality and integrity of the envelope content without cloning the sender. 

We remark that while this is a very strong adversary model, perhaps stronger than most practical attackers, and certainly stronger than the models discussed in some previous works \cite{cloneCounter, cloneSesame, cloneIdentityKey}, such an adversary could exist in practice and we believe their capabilities should be accounted for. One example of such an adversary could be a company that has installed CA certificates on their employees laptops to perform https deep-packet-inspection~\cite{10.1145/2829988.2787502}. Such an adversary would be able to bypass the protections afforded by TLS and also have an opportunity to clone the client.

\section{Protocol Description}
\label{sec:protocolDesc}

This section presents our changes to the Signal protocol as our novel solution. We first give the big picture on how the different protocols in our solution interact with one another for registration and session establishment. We then describe each protocol in detail where we represent each protocol in a message sequence chart, and our changes are highlighted in blue. The protocols described here utilise the algorithms introduced earlier in \cref{sec:background}.

\subsection{Overview}

\begin{figure}[t]
\centering
\includegraphics[scale = 1]{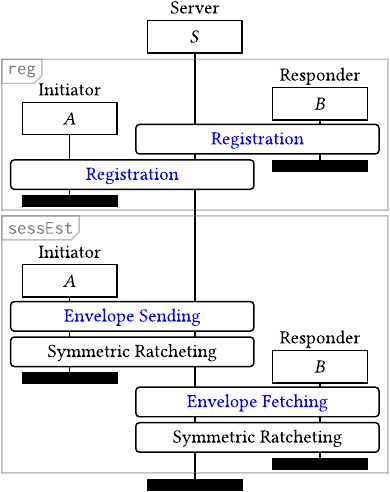}
\caption{An overview of the phases of registration ($\mathtt{reg}$) and session establishment ($\mathtt{sessionEst}$) integrated with our solution. Our modifications are applied in the Registration Protocol and the first envelope sent and received in the session establishment (and subsequent asymmetric ratcheting) where using the Envelope Sending Protocol and the Envelope Fetching protocol respectively. Symmetric ratcheting proceeds identically with the original Signal protocol. Recall from \cref{sec:sysAdvModel} that due to the asynchronicity of Signal, the communicating pair of clients are not required to be online at the same time.}
\label{fig:overview}
\end{figure}

An overview of our solution is shown in \cref{fig:overview} where our changes are applied at the start of the communication, i.e., in the phases of registration and session establishment. Note that the sequence of registration for the initiator and the responder is insignificant. The Envelope Sending Protocol and the Envelope Fetching Protocol are used to respectively send and receive the first envelope during session establishment (and the later asymmetric ratcheting). Recall from \cref{sec:background} that the established session is extended indefinitely by asymmetric ratcheting, i.e., when the initiator and the responder switches the role of the sender and the recipient of an envelope. The Envelope Sending Protocol and the Envelope Fetching Protocol are similarly applied for the first envelope during asymmetric ratcheting.


Notice that the phases of session establishment and asymmetric ratcheting in our solution is different from the original Signal Protocol. In our solution, the procedure for session establishment and asymmetric ratcheting are unified into one pair of sending and fetching protocols where both phases require the execution of the Envelope Sending Protocol and the Envelope Fetching Protocol. However, the protocol to run session establishment and asymmetric ratcheting in Signal are two different protocols since session establishment requires a client to first fetch the prekey bundle of its communicating partner from the server, but not for asymmetric ratcheting. 

As the symmetric ratcheting phase of the Signal protocol does not involve the exchange of new key materials, we do not require modification of the protocol in this phase. The symmetric ratcheting phase in our solution is identical with that in the original Signal protocol in which the sender sends subsequent envelopes to the server directly and the server then stores the envelopes until they are fetched by the recipient. 

\subsection{Registration Protocol}

\begin{figure}
\centering
\includegraphics[scale = 1]{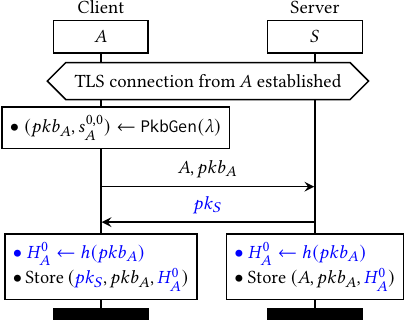}
\caption{The Registration Protocol where a client $A$ generates and sends its prekey bundle to the server $S$. The client and and the server both calculate and store the client's initial keychain computed from the generated public keys.}
\label{fig:reg}
\end{figure}

\cref{fig:reg} illustrates the Registration Protocol for a client $A$ to commit its prekey bundle to the server in the registration phase of the Signal protocol. The client executes the $\mathtt{PkbGen}$ algorithm to generate its prekey bundle $pkb_A$ and its initial state $s^{0, 0}_A$. After committing the prekey bundle to the server $S$, the server sends its long-term public key $pk_S$ for use in the later two protocols: the Envelope Sending Protocol and the Envelope Fetching Protocol. Both the client and the server then compute and store the initial \textit{keychain} $H^0_A$ of the client, which is a hash of the client's prekey bundle. The server stores the client's initial keychain with the client's prekey bundle whereas the client stores its prekey bundle, its initial keychain, and the newly received server's public key. The adversary is assumed to not be present in this protocol, and as per our adversary model, adversary $\mathcal{E}$ does not have write access to modify the newly received server's public key on the client's state but only read access when cloning the client. 

\subsection{Envelope Sending Protocol}
\label{sec:messSending}

\begin{figure}[!t]
\centering
\includegraphics[scale = 1]{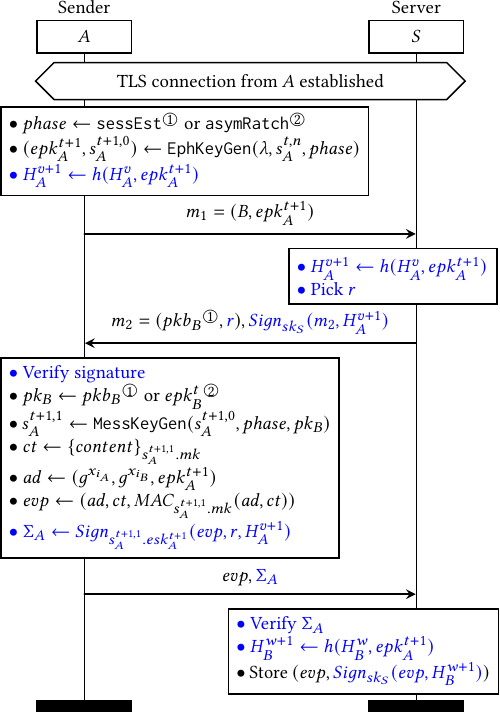}
\caption{The Envelope Sending Protocol in the session establishment phase and the asymmetric ratcheting phase of the Signal protocol. In this protocol, a sender $A$ sends the first envelope of the epoch intended for a recipient $B$ to the server. If the protocol is successful, the server accepts and stores the sender's envelope for the recipient. \textcircledcentred{1}: Parameters for session establishment. \textcircledcentred{2}: Parameters for asymmetric ratcheting.}
\label{fig:messSending}
\end{figure}

The Envelope Sending Protocol is depicted in \cref{fig:messSending} and the protocol is executed by a sender $A$ only when it sends the first envelope of a new epoch $t+1$ to a recipient $B$, in the two phases of the Signal protocol, namely the session establishment phase ($\mathtt{sessEst}$) and the asymmetric ratcheting phase ($\mathtt{asymRatch}$). After a successful run of this protocol, the sender sends the envelope intended for the recipient to the server $S$. The server buffers the envolope for the recipient until the recipient requests to fetch the envelope in the Envelope Fetching Protocol. 

The protocol starts with the establishment of TLS connection between the sender and the server, and we assume that the server has authenticated the sender on the TLS layer, e.g., through a username-password combination. The sender initiates the protocol by first executing the $\mathtt{EphKeyGen}$ algorithm, which generates fresh public keys $epk^{t+1}_A$ and the updated state $s^{t+1,0}_A$ of the sender for the new epoch $t+1$. The public keys are then hashed with the sender's current keychain $H^v_A$ to update the keychain to the new epoch $v+1$. Notice that we use $v$ to refer to the epoch of the keychain as the keychain is updated continuously with each epoch that the sender $A$ has participated across multiple communication sessions with multiple instances of recipients $B$ throughout the lifetime of the sender $A$, as opposed to $t$ which specifically refers to only the epochs in a single communication session between one particular pair of $A$ and $B$. The sender then sends message $m_1$ to the server, which comprises the recipient's identifier and the freshly generated public keys.

Once the server receives $m_1$, the server similarly updates its copy of the sender's current keychain $H^v_A$ using the received public keys $epk^{t+1}_A$ by hashing both information to generate the new keychain $H^{v+1}_A$. The server then issues a random nonce $r$ to the sender in message $m_2$, which additionally includes the prekey bundle $pkb_B$ of the recipient if the phase of the protocol execution is session establishment. The random nonce serves as a challenge for the sender to test if the sender indeed possesses the corresponding private keys $esk^{t+1}_A$ of $epk^{t+1}_A$. Message $m_2$ is sent together with the signature of the server on the message and the sender's keychain for the new epoch using the server's long-term private key $sk_S$ corresponding to the public key $pk_S$ sent in the Registration Protocol to detect the active tampering of adversary $\mathcal{E}$ (see \cref{gua:advDetect}). 

The sender first verifies the server's signature after receiving $m_2$. A successful verification assures the sender that the server has the same value of the sender's keychain implying that active Man-in-the-Middle (MitM) attacks do not take place against the sender up until the upcoming epoch $t + 1$. The sender proceeds to generate the ciphertext of the envelope's content for the recipient using the updated message key $mk^{t+1}_1$ contained in the updated state $s^{t+1,1}_A$ returned from running the $\mathtt{MessKeyGen}$ algorithm. If the phase is session establishment, then the public keys $pk_B$ of the recipient are $pkb_B$; otherwise $pk_B$ is the recipient's previous public ratchet key $epk^t_B$. The associated data $ad$ for the communication with the recipient consists of the long term identity public keys $g^{x_{i_A}}$ and $g^{x_{i_B}}$ of the sender and the recipient respectively, and $epk^{t+1}_A$. The sender then puts together the first envelope $evp$ for the recipient for the new epoch $t+1$, which consists of the associated data, the ciphertext, and a Message Authentication Code (MAC) under the updated message key $mk^{t+1}_1$ over both the associated data and the ciphertext. The sender subsequently generates a signature $\Sigma_A$ on the envelope using the fresh private keys $esk^{t+1}_A$ corresponding to $epk^{t+1}_1$. Lastly, the sender sends the envelope with $\Sigma_A$ to the server.

Upon receiving the envelope and the signature $\Sigma_A$, the server verifies $\Sigma_A$. If the verification passes, then the server has the assurance that the sender of the envelope is the same sender that initiates the protocol. The server then updates its copy of the recipient's current keychain $H^w_B$ to the new epoch $w+1$ by hashing $H^w_B$ with $epk^{t+1}_A$. Notice again that we use $w$ to indicate the epochs that the recipient has participated over its lifetime across multiple communication sessions with different instances of senders $A$, which might not necessarily equal to $v$, the epochs of the sender's keychain. The server generates a signature over the envelope and $H^{w+1}_B$, and stores the envelope and the signature until they are fetched by the recipient in the Envelope Fetching Protocol. 

\subsection{Envelope Fetching Protocol}

\begin{figure}
\centering
\includegraphics[scale = 1]{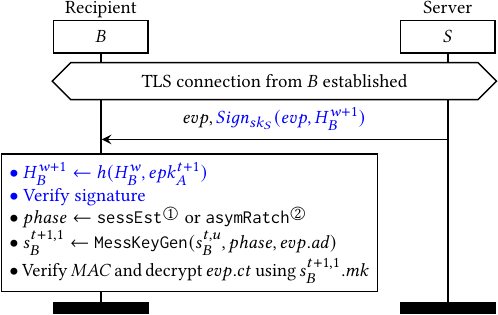}
\caption{The Envelope Fetching Protocol where the server $S$ delivers the buffered envelope from a sender $A$ to a recipient $B$ after the recipient has established a TLS connection with the server to fetch the envelope. \textcircledcentred{1}: Parameters for session establishment. \textcircledcentred{2}: Parameters for asymmetric ratcheting.}
\label{fig:messDelivery}
\end{figure}

As portrayed by \cref{fig:messDelivery}, the Envelope Fetching Protocol is for a recipient $B$ to fetch the envelope $evp$ sent by a sender $A$ which is buffered at the server $S$. The protocol starts with the recipient establishing a TLS connection to the server signalling that the recipient is now online and is requesting to fetch the envelope. Similarly, we assume that the recipient is authenticated by the server on the TLS level, e.g., through a username-password combination. Upon receiving the request, the server retrieves and sends the recipient the buffered envelope with the server's previously generated signature. After receiving the envelope, the recipient first calculates its keychain $H^{w+1}_B$ for the new epoch using $epk^{t+1}_A$ from the envelope. With this updated keychain, the recipient verifies the server's signature on the envelope and $H^{w+1}_B$. The verification of the server's signature assures the recipient that the sender of the envelope has indeed contacted the server before sending the envelope (see \cref{gua:evasionDetect}). The recipient then proceeds to execute the $\mathtt{MessKeyGen}$ algorithm to update its state to $s_B^{t+1,1}$ which contains the message key $mk$ for MAC verification and ciphertext decryption of the envelope to retrieve the content of the envelope. 

\section{Security Analysis}
\label{sec:secAnalysis}

We present the security analysis for our solution in this section. We informally prove that our solution provides new security guarantees in addition to retaining the existing security guarantees of Signal. This excludes deniability since deniability is hard to achieve in practice if the server already keeps a log of users, e.g., for logins. Each guarantee is associated with a specific goal of adversary $\mathcal{E}$ as described in \cref{sec:sysAdvModel}. Towards the end of this section we give a summary on how the individual guarantees complement one another to form a novel solution that detects the presence of a powerful active Man-in-the-Middle (MitM) adversary and limits the power of the adversary.

In the following proofs we use the notation $A'$ to indicate the adversary impersonating $A$, and $m_i$ to refer to the $i$-th protocol message.

\subsection{New Guarantees}

The two new security guarantees that our solution provides are the detection of active MitM adversary (\cref{gua:advDetect}) and the detection of non-compliance in protocol execution (\cref{gua:evasionDetect}). The Envelope Sending Protocol provides \cref{gua:advDetect} whereas the Envelope Fetching Protocol provides \cref{gua:evasionDetect}. \cref{gua:advDetect} and \cref{gua:evasionDetect} prevent Goal~(1) of the adversary as described in the adversary model in \cref{sec:sysAdvModel}. 

\begin{guarantee}[Envelope Sending Protocol: Active Adversary Detection]
\label{gua:advDetect}
When $A$ runs the Envelope Sending Protocol at epoch $t_A$, the presence of $\mathcal{E}$ is detected, if $\mathcal{E}$ has performed at least one run of the Envelope Sending Protocol impersonating $A$ between epochs $\tau$ and $t_A$.
\end{guarantee}

\begin{proof}
To break this guarantee, i.e., to remain undetected by both $A$ and $S$ simultaneously in the Envelope Sending Protocol, an active MitM adversary $\mathcal{E}$ has to generate a valid signature on both $m_2$, and $m_3$. 
As the adversary possesses $A$'s keychain $H^{\tau}_A$ from the epoch of compromise $\tau$, the adversary can compute a valid keychain $H^{t_{A'}}_{A'}$ to impersonate $A$, using $H^{\tau}_A$ and the public keys $epk^{t_{A'}+1}_{A'}$ freshly generated by the adversary for $m_1$. Note that during the epochs following $\tau$, $S$'s copy of $A$'s keychain is updated to $H^{t_{A'}}_{A'}$ while $A$'s copy of the keychain still remains at $H^{\tau}_A$. At epoch $t_A \ge t_{A'}$ when $A$ initiates the Envelope Sending Protocol, $A$ freshly generates $epk^{t_A}_A$ and a new keychain $H^{t_A}_A = h(H^{\tau}_A, epk^{t_A}_A)$ for sending $m_1$. Also note that at epoch $t_A$ since $A$ and $S$ each has a different copy of $A$'s keychain, the adversary is immediately detected by $A$ if the adversary does not tamper with $m_2$, or by $S$ if the adversary does not tamper with messages $m_1$ and $m_3$. 

For the adversary to be undetectable by $A$, this requires the adversary to be online at $t_A$ when $A$ runs the protocol for the adversary to change the signature of $S$ on $H^{t_A}_{A'}$ to $H^{t_A}_{A}$ instead. This then requires the adversary to possess the signing key $sk_S$ of $S$ and the adversary has only two options: either derive or guess $sk_S$. However, the former contradicts our assumption of secure cryptographic primitives and the latter can only occur with a negligible probability. 

If the adversary drops the last message containing the $evp$ sent by $A$ during $A$'s run of the Envelope Sending Protocol at epoch $t_A$, the adversary can send a new envelope $evp'$ with a content chosen by the adversary on behalf of $A$. However, now the adversary is required to generate the signature $\Sigma_A$ on $evp'$ and since $\Sigma_A$ is signed with $A$'s freshly generated $esk^{t_A}_A$ at epoch $t_A$, the adversary has three choices to generate $\Sigma_A$: (i)~use $esk^{t_A}_{A'}$ freshly generated by the adversary itself, (ii)~derive $esk^{t_A}_A$ from $epk^{t_A}_A$ sent by $A$ in $m_1$, or (iii)~simply guess $esk^{t_A}_A$. For the first case the adversary generates its own keypairs $(epk^{t_A}_{A'}, esk^{t_A}_{A'})$ at epoch $t_A$ and replaces the $epk^{t_A}_A$ from $A$ in $m_1$ with $epk^{t_A}_{A'}$ to generate a valid $\Sigma_{A'}$ on $evp'$. Since $epk^{t_A}_A$ is used in updating the keychain of $A$, this also requires the adversary to manipulate the keychain of $A$ and thus the signature of $S$ in $m_2$ to remain undetected. This is impossible by contradiction. For the second case, if the adversary successfully derives $A$'s freshly generated $esk^{t_A}_A$ from $epk^{t_A}_A$, then this breaks the Computational Diffie-Hellman (CDH) assumption since the generated keypairs are Diffie-Hellman keypairs and thus breaks our assumption of secure cryptographic primitives. For the last case the adversary can only successfully guess $A$'s $esk^{t_A}_A$ with a negligible probability due to the assumption of secure cryptographic primitives.
\end{proof}

\begin{guarantee}[Envelope Fetching Protocol: Non-Compliance Detection]
\label{gua:evasionDetect}
Every envelope fetched through the Envelope Fetching Protocol has been previously sent to the server through the Envelope Sending Protocol.
\end{guarantee}

\begin{proof}
Breaking this guarantee means that an adversary $\mathcal{E}$ evades detection by not executing the Envelope Sending Protocol with the server $S$, and instead sending an envolope $evp'$ of the adversary's choosing directly to $B$ in the Envelope Fetching Protocol while impersonating $S$. To achieve this while also remaining undetectable from $B$, the adversary must produce the signature of $S$ on $evp'$, which requires the adversary to have the private key $sk_S$ of $S$. As we have seen in \cref{gua:advDetect}, the adversary can only achieve this with two options: either derive or guess $sk_S$. The former violates our assumption of secure cryptographic primitives and the latter can only occur negligibly.
\end{proof}

\subsection{Existing Guarantees}

In addition to providing new guarantees, both protocols in our solution maintain the existing properties from the original Signal protocol. These are, perfect forward secrecy (\cref{gua:pfs}), post compromise security (\cref{gua:pcs}), and end-to-end encryption (\cref{gua:confInt}). With the adversary goals described in the adversary model in \cref{sec:sysAdvModel}, \cref{gua:pfs} and \cref{gua:pcs} prevent Goal~(2) and Goal~(3) of the adversary respectively whereas \cref{gua:confInt} prevents Goal~(4). 

\begin{guarantee}[All Protocols: Perfect Forward Secrecy]
\label{gua:pfs}
Adversary $\mathcal{E}$ is unable to reveal the content of past envelopes from epochs $t < \tau$. 
\end{guarantee}

\begin{proof}
For the adversary to break this guarantee, the adversary has to derive the root key of previous epochs $t < \tau$, for which the adversary has two options: (1)~using messages exchanged between a sender $A$ and a recipient $B$ from previous runs of the Envelope Sending Protocol and Envelope Fetching Protocol during epochs $t < \tau$, in addition to (2)~using the secrets cloned from $A$ at epoch $\tau$. (1)~In the first option, since only the public keys $epk^t_A$ are exchanged in all three messages in the Envelope Sending Protocol during epochs $t < \tau$, the adversary has to either derive the corresponding private keys $esk^t_A$ from $epk^t_A$ or simply guess $esk^t_A$. The former breaks the CDH assumption whereas the probability of the latter occurring is negligible. (2)~If the goal of the adversary is only the root key $rk^{\tau - 1}$ of epoch $\tau - 1$, then the adversary requires not only the private ratchet key $x ^ {\tau - 1}_A$ of $A$ at epoch $\tau - 1$ but also the root key $rk ^ {\tau - 2}$ from epoch $\tau - 2$. As the adversary obtains only $rk^{\tau}$ and $x_A^{\tau}$ from cloning $A$ at epoch $\tau$, the adversary is in turn left with only two choices: (2a)~finding the inverse of the key derivation function to obtain $rk^{\tau - 1}$ from $rk^{\tau}$, or (2b)~brute force $rk^{\tau - 1}$. In successfully achieving the former the adversary breaks the assumption of a secure key derivation function and thus the assumption of secure cryptographic primitives whereas the latter can only occur with a negligible probability.
\end{proof}

\begin{guarantee}[All Protocols: Post Compromise Security]
\label{gua:pcs}
If a sender $A$ sends an envelope to a recipient $B$ at an epoch $\tau + 1$ and the adversary $\mathcal{E}$ is passive at both $\tau + 1$ and $\tau + 2$, then the content of future envelopes in epochs $t \ge \tau + 2$ is only known to $A$ and $B$, and $\mathcal{E}$ is since locked out of future communication.
\end{guarantee}

\begin{proof}
To break this guarantee, the adversary has to be able to reveal the content of future envelopes from epochs $t \ge \tau + 2$ and this requires the adversary to be able to generate the root key $rk^t$ in epochs $t \ge \tau + 2$. This means that the adversary must be able to first compute the root key $rk^{\tau + 2}$ of the epoch $\tau + 2$ before the adversary can compute the root key of subsequent epochs $t > \tau + 2$. While the adversary can still generate the root key $rk^{\tau + 1}$ when $B$ switches role to become the sender at epoch $\tau + 1$ and thus reveal the content of the envelopes sent by $B$ in that epoch, the adversary still requires the fresh private ratchet key $esk^{\tau + 2}_A$ generated by $A$ at the next epoch $\tau + 2$ in order to generate $rk^{\tau + 2}$. Note that at $\tau + 2$ when $A$ switches back to be the sender, $A$ performs asymmetric ratcheting by generating a fresh ratchet keypair $(epk^{\tau + 2}_A, esk^{\tau + 2}_A)$. Since the adversary is passive when $A$ sends the public ratchet key $epk^{\tau + 2}_A$ in $m_1$ of the Envelope Sending Protocol, i.e., the adversary does not interfere with the message by injecting its own ratchet keypair, to compute $rk^{t_A + 2}$, the adversary must obtain the private ratchet key $esk^{\tau + 2}_A$ corresponding to the $epk^{\tau + 2}_A$ sent by $A$. This leaves the adversary with only two options: to derive $esk^{\tau + 2}_A$ from $epk^{\tau + 2}_A$, or to guess $esk^{\tau + 2}_A$. The former again breaks the CDH assumption and thus contradicts our assumption of secure cryptographic primitives whereas the latter can only happen with a negligible probability.
\end{proof}

\begin{guarantee}[All Protocols: End-to-End Content Confidentiality and Integrity]
\label{gua:confInt}
In any epoch $t$, the content of an envelope is only known to the sender $A$ and the recipient $B$ of the envelope, and the content received by $B$ is the same content as sent by $A$. An adversary $\mathcal{E}$ can reveal and change the content sent by $A$ only if $\mathcal{E}$ uses its cloning ability. 
\end{guarantee}

\begin{proof}
In epochs $t \ge 1$, i.e., epochs after session establishment, for the adversary to break this guarantee, the adversary should be able to decrypt or change the content of the envelope without using its cloning ability. The adversary can either (1) inject the adversary's own chosen envelope in either the Envelope Sending Protocol or the Envelope Fetching Protocol, or (2) generate the root key $rk^t$ of a epoch $t$ that the adversary chooses.

(1)~The adversary can choose to inject an envelope $evp'$ of its own choosing in either (1a)~the last message of the Envelope Sending Protocol, or (1b)~the message sent to $B$ in the Envelope Fetching Protocol. (1a)~For the adversary to inject $evp'$ in the last message of Envelope Sending Protocol, the adversary has to generate a valid signature $\Sigma_A$ on $evp'$. As $\Sigma_A$ is generated from $A$'s fresh private keys $esk^t_A$, the options that the adversary has are to: either derive $esk^t_A$ from $A$'s freshly generated public keys $epk^t_A$ or simply guess $esk^t_A$. (1b)~Similarly for the adversary to inject $evp'$ into the Envelope Fetching Protocol, the adversary has to generate a valid signature on $evp'$ from the server. As we have established before, this again means that the adversary has two options: derive the private key $sk_S$ of the server $S$, or guess $sk_S$. 

(2)~To derive the root key $rk^t$ of any epoch $t$ without having access to the secrets of a sender $A$ via cloning, the adversary has to compute the root key $rk^{t - 1}$ at epoch $t - 1$, which itself is derived based on the root key $rk^{t - 2}$ at epoch $t - 2$, and this goes on until the first root key $rk^1$ established in the session establishment phase. This ultimately means that the adversary has to first be able to derive $rk^1$ and there are only three ways for the adversary to do so. (2a)~The adversary at epoch $t=1$ generates and inserts its own prekey bundle $pkb_{A'}$ in $m_2$ of the Envelope Sending Protocol sent by $A$. This again requires the adversary to be able to generate $S$'s signature, and as we have seen many times, there are only two ways for the adversary to achieve this: deriving or guessing the private key $sk_S$ of $S$. (2b)~The adversary chooses to derive or guess the corresponding private keys $esk^t_A$ of $A$ from $A$'s freshly generated public keys $epk^t_A$ sent in $m_1$ of the Envelope Sending Protocol, or the private keys of $B$ contained in $B$'s initial state $s^{0,0}_B$ corresponding to the public keys in $B$'s prekey bundle sent by the server in $m_2$ of the Envelope Sending Protocol. (2c) Lastly the adversary simply guesses $rk^1$. 

In all cases, deriving the private key of $A$, $S$, or $B$ can only occur with negligible probability.
\end{proof}

\subsection{Summary}

To summarise, \cref{gua:confInt} assures that the adversary $\mathcal{E}$ is unable to reveal the content of the envelopes without using its cloning ability. If the adversary does use its cloning ability, from \cref{gua:pfs} and \cref{gua:pcs}, the adversary can reveal the content of the envelopes in only two epochs, $\tau$ and $\tau + 1$, with the condition that $\mathcal{E}$ is passive in both epochs. Otherwise, if the adversary is active in epochs between $\tau$ and $t_{A}$, then from \cref{gua:advDetect} and \cref{gua:pfs} the adversary can only inject its own envelopes in these epochs until the adversary is detected at epoch $t_A$ when the cloned sender $A$ runs the Envelope Sending Protocol. All three guarantees, \cref{gua:advDetect}, \cref{gua:pfs}, \cref{gua:pcs}, essentially provide the adversary using its cloning ability two choices: either risk being locked out of the communication, or risk being detected. Finally, \cref{gua:evasionDetect} ensures that all envelopes, including those sent by an active adversary, go through both the Envelope Sending Protocol and the Envelope Fetching Protocol before they reach their intended recipient.

\section{Implementation and Evaluation}
\label{sec:implEval}

To demonstrate the practicality of our solution, we implemented a proof-of-concept that builds on the open-source Signal protocol library \cite{signalImpl}. This section presents the findings and evaluations on the  implementation that forms the basis of our experiment.

\subsection{Implementation Details}

\begin{figure*}[!t]
\centering
\begin{subfigure}[t]{0.47\linewidth}
\centering
\includegraphics[width=\linewidth]{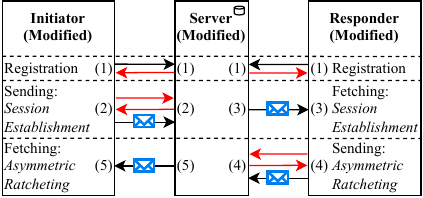}
\caption{Implementation for our modified Signal protocol.}
\label{fig:expSetup_mod}
\end{subfigure}
\hspace*{0.25cm}
\begin{subfigure}[t]{0.47\linewidth}
\centering
\includegraphics[width=\linewidth]{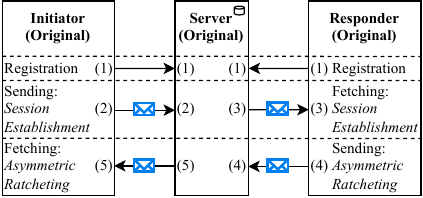}
\caption{Implementation for the original Signal protocol.}
\label{fig:expSetup_ori}
\end{subfigure}
\caption{An overview of the two versions of our implementation for our modified protocols and the original Signal protocol. In both versions of the implementation, each client performs three actions (registering, sending, fetching) in the three different Signal communication phases (registration, session establishment, asymmetric ratcheting) with the server, who manages a database for storage. The arrows indicate the number of exchanged protocol messages in the respective version of the implementation until an envelope is transferred, and the red arrows indicate the extra interaction between the client and the server in our modified protocols, which is the focus of the experiment.}
\label{fig:expSetup}
\end{figure*}

Our implementation is built on top of the open-source Rust Signal library (version \texttt{0.32.1}) \cite{signalImpl}. Our modifications on the Signal protocol integrate seamlessly with the library, and we use the cryptographic primitives already available in the library. We make our implementation available on our git server\footnote{\textit{As the URL to the repository deanonymises the authors, the URL will only be included in the camera-ready version of the paper.}} and only describe its high-level details here. 

\cref{fig:expSetup} depicts an overview of the two versions of our implementation. \cref{fig:expSetup_mod} depicts the implementation on our modified Signal protocol and \cref{fig:expSetup_ori} depicts the implementation on the original Signal protocol. To test the performance of the three different actions of a client throughout the three communication phases of Signal where we have applied our changes, we design the implementation in a way that a client, an initiator and a responder, in both versions goes through, in sequnce: (1) registration with the server, (2) sending an envelope to the server, and (3) fetching an envelope from the server. Recall that for an initiator sending an envelope first occurs at session establishment whereas fetching an envelope first occurs at asymmetric ratcheting; for the responder the sending first occurs at asymmetric ratcheting whereas the fetching first occurs at session establishment. 

Observe that the arrows for (1), (2), and (3) in \cref{fig:expSetup_mod} represent the same number of messages exchanged in the Registration Protocol (\cref{fig:reg}), the Envelope Sending Protocol (\cref{fig:messSending}), and the Envelope Fetching Protocol (\cref{fig:messDelivery}) respectively. Notice that the difference between \cref{fig:expSetup_mod} and \cref{fig:expSetup_ori} is the number of messages exchanged between a client and the server in the protocol until an envelope is sent or received by the client. Also notice that the server in \cref{fig:expSetup_ori} only stores and relays the envelopes between clients without performing any other action in the protocol.

\subsection{Experimental Setup}

We investigate the performance overhead of the discrepancy in message processing between both versions of implementation in \cref{fig:expSetup} experimentally. Specifically our experiment measures the execution time for the protocols in \cref{fig:expSetup}, i.e., (1) to (5), to be completed by the client and the server in both versions of the implementation where we record the protocol completion time on the client's end and the server's end. We conduct the experiment with $n$ instances of initiators and $n$ instances of responders executing each protocol with a single instance of the server in both versions of the implementation. To make network latency negligible, we execute all client instances and the server on a single machine running Ubuntu~22.04 with Intel\textregistered{} Core\textsuperscript{TM} i5-1145G7. 

\subsection{Results Analysis}

\begin{figure*}[!t]%
\centering%
\begin{subfigure}[T]{0.48\linewidth}%
\centering
\includegraphics[width=\linewidth]{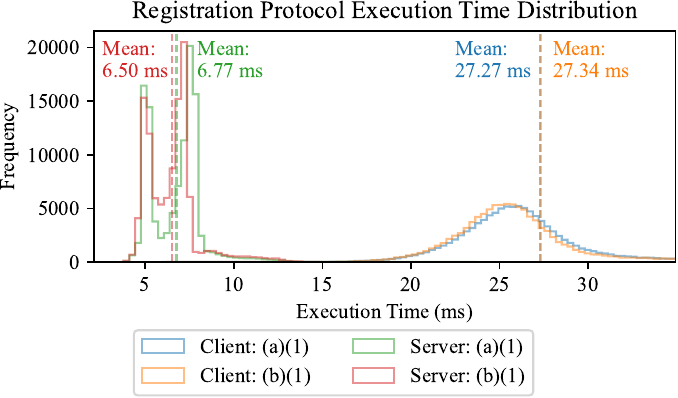}%
\caption{The distribution of the execution time of $n = 110000$ clients ($n = 55000$ initiators and $n = 55000$ responders) for the Registration Protocol with the server.}%
\label{fig:regDist}%
\end{subfigure}\hfill%
%
\begin{subfigure}[T]{0.48\linewidth}%
\centering%
\includegraphics[width=\linewidth]{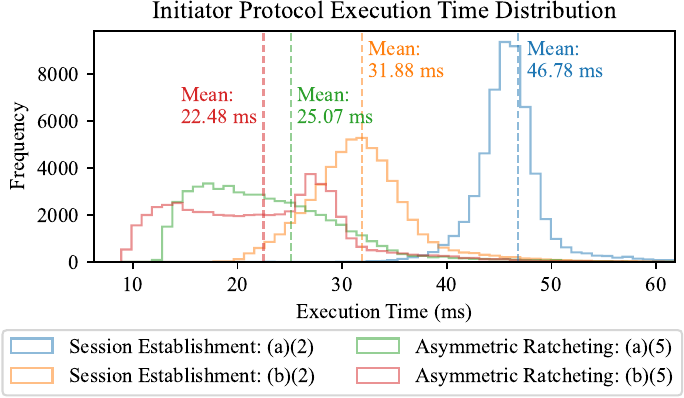}%
\caption{The distribution of the execution time of $n = 55000$ initiators for session establishment (sending) and asymmetric ratcheting (fetching).}%
\label{fig:aliceDist}%
\end{subfigure}%

\begin{subfigure}[T]{0.48\linewidth}%
\centering%
\includegraphics[width=\linewidth]{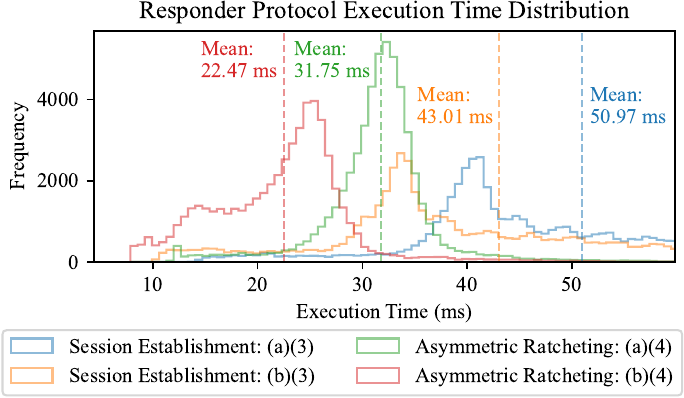}%
\caption{The distribution of the execution time of $n = 55000$ responders for session establishment (fetching) and asymmetric ratcheting (sending).}%
\label{fig:bobDist}
\end{subfigure}\hfill%
%
\begin{subfigure}[T]{0.48\linewidth}%
\centering%
\includegraphics[width=\linewidth]{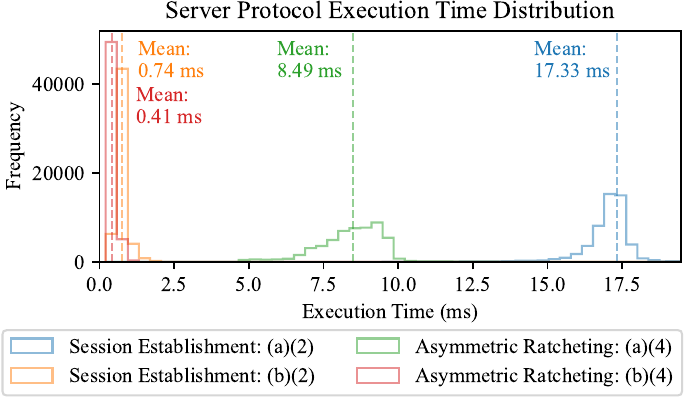}%
\caption{The distribution of the execution time for the server to complete $n = 55000$ requests for session establishment from the initiators and $n = 55000$ requests for asymmetric ratcheting from the responders, as well as a total of $n = 110000$ requests of fetching from all clients.}%
\label{fig:serverDist}%
\end{subfigure}%

\caption{Distributions of execution time for each action performed by the clients (both initiators and responders) and the server in the two versions of the implementation. (a) and (b) refer to \cref{fig:expSetup_mod} and \cref{fig:expSetup_ori} respectively whereas (1) to (6) refer to the actions (registration, sending, and fetching) taken by the clients with the server in the two subfigures.}%
\label{fig:dist}%
\end{figure*}

We run an experiment with $n = 55000$ instances of initiators and $n = 55000$ instances of responders with a single instance of the server in both versions of the implementation. The results representing the distribution of the protocol execution time are visualised in \cref{fig:dist} as histograms. We calculate the difference in mean as the performance overhead in each specific protocol run.

The results in \cref{fig:regDist} show the distribution of the protocol execution time for the registration of both clients (initiators and responders) with the server. The difference in mean between the two implementation versions for the execution time is 0.06 ms on the clients' ends and 0.27 ms on the server's end. Both differences in mean are less than 0.5 ms and they show that our Registration Protocol has an insignificant performance overhead than the original Signal's Registration Protocol. This means that a client and the server experience no difference when using our Registration Protocol in terms of processing the messages during registration.

\cref{fig:aliceDist} depicts the distribution of the protocol execution time for the initiators to send an envelope to the server at session establishment and to receive an envelope from the server at asymmetric ratcheting in both versions of the implementation. At session establishment, the difference in mean is only 14.90 ms for sending an envelope to the server, and at asymmetric ratcheting, fetching an envelope from the server observes a difference in mean of only 2.59 ms. Similarly, \cref{fig:bobDist} illustrates the distribution of the protocol execution time for the responders to receive an envelope from the server at session establishment and to send an envelope to the server at asymmetric ratcheting in both versions of the implementation. Fetching an envelope at session establishment sees a difference in mean of just 7.97 ms whereas sending an envelope at asymmetric ratcheting sees a difference in mean of just 9.28 ms. As the differences in mean are all less than 15~ms, this demonstrates that there is hardly a noticeable difference on the client-side processing using our protocols. Thus our changes are applicable for real-world deployment without deteriorating the user experience on the clients.

\cref{fig:serverDist} illustrates the distribution of the protocol execution time for completing the Envelope Sending Protocol on the server's end. The difference in mean execution time at session establishment is 8.08 ms whereas at asymmetric ratcheting it is 16.59 ms. In all cases, the differences in mean are still well under 20 ms. This again demonstrates the negligible performance overhead of our modified protocols for the server to handle the requests from the clients. Our modified protocols thus do not affect the usability of the original Signal protocol since a human user can barely notice the 20 ms difference when waiting for the server to complete the sending or the receiving of a message.

Overall, the experiment results indicate that the performance overhead for our protocols is a trivial constant and from that, our solution exhibits a linear time complexity with the number of clients in the system. We can thus conclude that our solution is on the same order as the performance of the original Signal protocol. This is an acceptable trade-off for the novel and stronger security guarantees that our solution provides. Additionally the performance of our solution can still be improved once it is deployed on a real-world application that has a more powerful infrastructure.

\section{Discussion}

While our adversary model specifies that the adversary is assumed to have only a one-time use of its cloning ability to get access to a device's secrets, this does not mean that the strong adversary detection mechanism of our solution do not hold if the adversary can clone a device more than once throughout the long-lived communication session. In fact, our adversary detection guarantee (\cref{gua:advDetect}) still holds even if an adversary is more powerful than our adversary model and has the ability to clone a device more than one time. The only condition here is that the communication channel has been `healed' (adversary locked out) before each cloning. Recall from \cref{gua:advDetect} (adversary detection) and \cref{gua:pcs} (post-compromise security) that after cloning, the adversary will either be locked out of the communication if it is passive, or the adversary will be detected if it is active. Thus, if a stronger adversary chooses to be locked out of the communication to avoid detection and uses its cloning ability again at a later epoch, this effectively limits the epochs of which the adversary can reveal the content of the envelopes in these epochs due to the self-healing property of the Signal protocol. So for the stronger adversary to perpetually remain on the communication the adversary has to be active on the channel, which will then trigger detection. Our solution thus makes it possible for this process to repeat with ongoing detection and not requiring any user participation, which coincides with the motivation of our work.

\section{Conclusion}

The emerging threat of breaking end-to-end encryption to get access to a user's private user data including private instant messaging conversations due to government coercion has the potential to proliferate Man-in-the-Middle (MitM) attacks. Current solutions \cite{signalSafetyNumber, signalSesame, euroSP, cloneIdentityKey} in the Signal protocol for key authentication to detect MitM attacks depend on a communicating pair of users to compare the fingerprints of their keys in an out-of-band manner. However, the issues with relying on user participation to ensure key authentication present usability difficulties as several studies have shown \cite{usabilityBlog, usabilityUsenix1, usabilityUsenix2, usabilityNdss}. Furthermore, this so-called authentication ceremony \cite{authCeremony} is expected to carry out every so often \cite{euroSP,cloneIdentityKey} and this again presents practicality issues.

We build our novel solution for these problems directly into the existing Signal protocol without requiring any additional entities. Our solution allows key authentication to be done in-band with the same communication channel used to send and receive messages between the communicating parties, without being dependent of user intervention. We achieve this by making the most of the Signal server, who is always present on the communication channels, and in practice who is expected to be honest in forwarding messages and distributing `prekey bundles' of the users. We assume the presence of a powerful adversary that is not only capable of performing active MitM attacks, but also gaining one-time access to all secrets on one of the two communicating parties. By embedding a recursively updating key fingerprint of a client as part of the client's identity, our solution allows both the server and the client to keep track of the changes in the client's key fingerprint, thus automating the detection for the presence of such an adversary. Our security analysis has shown that not only does our solution provide new security guarantees, but does so in such a way that preserves the underlying end-to-end encryption while also keeping the existing properties of the Signal protocol (except deniability).

We implemented our proof-of-concept of the solution based on the open-source Signal protocol library, which is the underlying library used by the actual messaging applications, and we use this to demonstrate the practicality and integrability of our solution into the library. Based on the implementation we run an experiment to determine the feasibility of the performance overhead of our solution as compared to the original Signal protocol. Our experimental results have shown that the performance for our solution has a linear overhead and is thus on the same order as the original Signal protocol.


\bibliographystyle{ACM-Reference-Format}
\bibliography{ref}

\appendix

\section{Algorithms}
\label{app:algorithms}

We explain the details of the algorithms that generate or update the session keys in each of the four phases of the Signal protocol here, i.e., the registration phase, the session establishment phase, the asymmetric ratcheting phase, and the symmetric ratcheting phase. The generation and the update of the session keys require three cryptographic primitives: a Diffie-Hellman key generation function and three key derivation functions (KDFs). The Diffie-Hellman key generation function is denoted $(g^x, x) \leftarrow DHGen(\lambda)$ which takes in a security parameter $\lambda$ and outputs a Diffie-Hellman keypair. The first KDF is denoted $k_1 \leftarrow KDF_1(x_1, x_2, x_3, x_4)$ that takes in four key materials $x_1, x_2, x_3, x_4$, and outputs one symmetric key $k_1$. The second KDF is denoted $(k_1, k_2) \leftarrow KDF_2(x_1, x_2)$ that takes in two key materials $x_1, x_2$ and outputs two symmetric keys $k_1, k_2$. Lastly, the third KDF is denoted $(k_1, k_2) \leftarrow KDF_3(x_1)$ that takes in only one key material $x_1$ but also outputs two symmetric keys $(k_1, k_2)$.

\subsection{Prekey Bundle Generation}

\begin{algorithm}[!t]
\SetAlgoVlined
\DontPrintSemicolon
\SetKwFunction{pkbGen}{$\mathtt{PkbGen}$}
\SetKwProg{prod}{Function}{}{}

\prod{\pkbGen{$\lambda$}}{
  \nl $(g^{x_{i}}, x_{i}) \leftarrow DHGen(1^\lambda)$ \;
  \nl $(g^{x_{s}}, x_{s}) \leftarrow DHGen(1^\lambda)$ \;
  \nl $(g^{x_{o}}, x_{o}) \leftarrow DHGen(1^\lambda)$ \;
  \nl $\sigma \leftarrow Sign_{x_{i}}(g^{x_{s}})$ \;
  \nl $pkb \leftarrow (g^{x_{i}}, g^{x_{s}}, g^{x_{o}}, \sigma)$ \;
  \nl $s^{0,0} \leftarrow (x_{i}, x_{s}, x_{o})$ \;
  \nl \KwRet{$(pkb, s^{0,0})$} \;
}{}
\caption{Generation of prekey bundle for a client.}
\label{alg:pkbGen}
\end{algorithm}

\noindent The Signal protocol starts with the registration phase where a client generates and registers its prekey bundle to the server. The generation of a client's prekey bundle follows \cref{alg:pkbGen}. The algorithm first generates three Diffie-Hellman keypairs, known respectively as the identity keypair $(g^{x_{i}}, x_{i})$, the signed prekey keypair $(g^{x_{s}}, x_{s})$, and the one-time prekey keypair $(g^{x_{o}}, x_{o})$. The algorithm then generates a signature on the public signed prekey using the private identity key. The prekey bundle $pkb$ consists of the three generated public keys and the generated signature. The initial state $s^{0,0}$ of the client consists of the three corresponding private keys. Both the prekey bundle and the initial state are the outputs returned by the algorithm.

\subsection{Ephemeral Keys Generation}

\begin{algorithm}[!t]
\SetAlgoVlined
\DontPrintSemicolon
\SetKwFunction{ephKeyGen}{$\mathtt{EphKeyGen}$}
\SetKwProg{prod}{Function}{}{}

\prod{\ephKeyGen{$\lambda$, $s^{t,n}$, phase}}{
  \If{$phase = \mathtt{sessEst}$}{
    \nl $(g^{x_{b}}, x_{b}) \leftarrow DHGen(1 ^ \lambda)$ \;
    \nl $(g^{x^1}, x^1) \leftarrow DHGen(1 ^ \lambda)$ \;
    \nl $epk^{t+1} \leftarrow (g^{x_{b}}, g^{x^1})$ \;
    \nl $esk^{t+1} \leftarrow (x_{b}, x^1)$ \;
  }
  \ElseIf{$phase = \mathtt{asymRatch}$}{
    \nl $(g^{x^{t+1}}, x^{t+1}) \leftarrow DHGen(1 ^ \lambda)$ \;
    \nl $epk^{t+1} \leftarrow (g^{x^{t+1}})$ \;
    \nl $esk^{t+1} \leftarrow (x^{t+1})$ \;
  }
  \nl $s^{t+1,0} \leftarrow s^{t,n}_A \cup esk^{t+1}$ \;
  \nl \KwRet{$(epk^{t+1}, s^{t+1,0})$} \;
}{}
\caption{Ephemeral Diffie-Hellman keypairs generation for a sender in session establishment or asymmetric ratcheting as required by the Triple Diffie-Hellman key agreement algorithm or the Double Ratchet algorithm respectively.}
\label{alg:ephKeyGen}
\end{algorithm}

\noindent Before a sender establishes or updates session keys in the phase of session establishment or in asymmetric ratcheting respectively, the sender first generates new Diffie-Hellman keypairs as described in \cref{alg:ephKeyGen}. For session establishment ($\mathtt{sessEst}$), the algorithm generates two keypairs: a base keypair $(g^{x_{b}}, x_{b})$, and the first ratchet keypair $(g^{x^1}, x^1)$. Similarly in asymmetric ratcheting ($\mathtt{asymRatch}$), the algorithm generates a single ratchet keypair $(g^{x^{t+1}}, x^{t+1})$ for the new epoch $t+1$. The algorithm then collects the freshly generated public keys into a tuple $epk^{t+1}$ and private keys into a tuple $esk^{t+1}$. The current state $s^{t,n}$ of the sender is then updated to $s^{t+1,0}$ that includes the freshly generated private keys. The algorithm then returns the tuple $epk^{t+1}$ and the updated state of the sender as outputs. As stated in \cref{sec:background}, the generation of new keypairs is not required for a recipient of an envelope in session establishment or asymmetric ratcheting.

\subsection{Message Key Generation}

\begin{algorithm}[!t]
\SetAlgoVlined
\DontPrintSemicolon
\SetKwFunction{messKeyGen}{$\mathtt{MessKeyGen}$}
\SetKwProg{prod}{Function}{}{}

\prod{\messKeyGen{$s^{t,n}$, phase, $pk_B$}}{
  \If{$phase = \mathtt{sessEst}$}{
    \nl Sender: $(x_{i}, x_{b}, x^1) \leftarrow s^{t,n}$ \\ Recipient: $(x_{i_B}, x_{s_B}, x_{o_B}) \leftarrow s^{t,n}$\;
    \nl Sender: $(g^{x_{i_B}}, g^{x_{s_B}}, g^{x_{o_B}}) \leftarrow pk_B$ \\ Recipient: $(g^{x_{i}}, g^{x_{b}}, g^{x^1}) \leftarrow pk_B$ \; 
    \nl $ms \leftarrow KDF_1(g^{x_{s_B}x_{i}}, g^{x_{i_B}x_{b}}, g^{x_{s_B}x_{b}}, g^{x_{o_B}x_{b}})$\;
    \nl $(rk^1, ck^1_0) \leftarrow KDF_2(ms, g^{x_{s_B}x^1})$ \;
    \nl $(ck^1_1, mk^1_1) \leftarrow KDF_3(ck^1_0)$ \;
    \nl $(rk, ck, mk) \leftarrow (rk^1, ck^1_1, mk^1_1)$ \;
    \nl Sender: $s \leftarrow s^{t,n} \setminus (x_{o}, x_{b}, x^1)$ \\ Recipient: $s \leftarrow s^{t,n} \setminus (x_{o_B})$ 
  }

  \ElseIf{$phase = \mathtt{asymRatch}$}{
    \nl Sender: $(x^{t+1}, rk^t) \leftarrow s^{t,n}$ \\ Recipient: $(x^t, rk^t) \leftarrow s^{t,n}$ \;
    \nl Sender: $(g^{x^t}) \leftarrow pk_B$ \\ Recipient: $(g^{x^{t+1}}) \leftarrow pk_B$ \;
    \nl $(rk^{t+1}, ck^{t+1}_0) \leftarrow KDF_2(rk^t, g^{x^tx^{t+1}})$ \;
    \nl $(ck^{t+1}_1, mk^{t+1}_1) \leftarrow KDF_3(ck^{t+1}_0)$ \;
    \nl $(rk, ck, mk) \leftarrow (rk^{t+1}, ck^{t+1}_1, mk^{t+1}_1)$ \;
    \nl $s \leftarrow s^{t,n} \setminus (rk^t, ck^t_u, mk^t_u)$ \;
  }

  \ElseIf{$phase = \mathtt{symRatch}$}{
    \nl $(rk^t, ck^t_u, mk^t_u) \leftarrow s^{t,n}$ \;
    \nl $(ck^t_{n+1}, mk^t_{n+1}) \leftarrow KDF_3(ck^t_u)$ \;
    \nl $(rk, ck, mk) \leftarrow (rk^t, ck^t_{n+1}, mk^t_{n+1})$ \;
    \nl $s \leftarrow s^{t,n} \setminus (rk^t, ck^t_u, mk^t_u)$ \;
  }
  \nl $s^{t,n+1} \leftarrow s \cup (rk, ck, mk)$ \;
  \nl \KwRet{$s^{t,n+1}$} 
}{}
\caption{Message key generation based on the the Triple Diffie-Hellman key agreement algorithm for the session establishment phase, or the Double Ratchet algorithm for the asymmetric ratcheting phase, as well as the symmetric ratcheting phase for a client and its communicating partner $B$ according to both clients' role in each phase.}
\label{alg:messKeyGen}
\end{algorithm}

The Diffie-Hellman keypairs generated in \cref{alg:pkbGen} for a sender are then used to establish or update the session keys in \cref{alg:messKeyGen} in session establishment or asymmetric ratcheting respectively to generate the envelope. As the secrets stored in the state of a sender is different from a recipient, some steps in the algorithm are executed differently depending on whether the client running the algorithm is a sender or a recipient but the output of the symmetric session keys are identical for both clients. For these steps we will specify clearly the operations that are taken respectively by a sender or a recipient. If a step is not labelled whether it is taken by a sender or a recipient, both the sender and the recipient execute the same operation. The algorithm is divided into three parts depending on the phase: the session establishment ($\mathtt{sessEst}$) phase, the asymmetric ratcheting ($\mathtt{asymRatch}$) phase, and the symmetric ratcheting ($\mathtt{symRatch}$) phase.

For session establishment, a sender executing this algorithm first retrieves its private identity key $x_{i}$, private base key $x_{b}$, and the first private ratchet key $x^1$ from its state. The sender also retrieves its communicating partner's, i.e., the recipient $B$'s public identity key $g^{x_{i_B}}$, public signed prekey $g^{x_{s_B}}$, and public one-time prekey $g^{x_{o_B}}$ from the recipient's public keys. Conversely, a recipient $B$ executing this algorithm first extracts its private identity key $x_{i_B}$, private signed prekey $x_{s_B}$, and the private one-time prekey $x_{o_B}$ from its state. The recipient also extracts, its communicating partner's, i.e., the sender's public identity key $g^{x_{i}}$, public signed prekey $g^{x_{b}}$, and the first public ratchet key $g^{x^1}$ from the sender's public keys. Both the sender and the recipient then generate the initial session keys of the communication session based on the Triple Diffie-Hellman (X3DH). The three session keys are the root key $rk$, the chain key $ck$, and the message key $mk$. The chain key and the message key are derived from the root key, and the message key is used to encrypt the content of an envelope. After the session keys are generated, the sender then removes its private one-time prekey $x_{o}$, private base key $x_{b}$, and the first private ratchet key $x^1$ from its state. The recipient similarly removes its private one-time prekey $x_{o_B}$ from its state.

In the asymmetric ratcheting phase, for a sender executing this algorithm, the sender first retrieves the private ratchet key $x^{t+1}$ for the new epoch $t+1$ as well as the root key $rk^t$ for the current epoch $t$ from its state. The sender additionally retrieves its communicating partner's, i.e., the recipient $B$'s public ratchet key $g^{x^t}$ for the current epoch from the recipient's public keys. In contrast, a recipient executing this algorithm first extracts its private ratchet key $x^t$ and the root key $rk^t$ for the current epoch from its state. This is followed by the recipient extracting its communicating partner's, i.e., sender $A$'s, public ratchet key $g^{x^{t+1}}$ for the new epoch. Both the sender and the recipient then generate the session keys for the new epoch $t+1$ based on the Double Ratchet algorithm, after which they subsequently remove the three session keys of the current epoch $t$ from their respective state. 

For symmetric ratcheting, no new key material is introduced. A sender and a recipient both derive the new session keys the same way in this phase according to the Double Ratchet algorithm. The sender and the recipient derive the next chain key and message key for the new envelope $n+1$ by passing the chain key for the current envelope $n$ into the KDF. The respective state of the sender and the recipient is then updated by removing the chain key and the message key for the current envelope $n$. 

As a last step, the newly established or updated session keys are added into the state of both the sender and the recipient. The algorithm returns the updated state containing the new session keys as output.

\end{document}